\documentclass[12pt]{iopart}
\pdfoutput=1
\usepackage{graphicx}
\usepackage{iopams}
\usepackage{cite}
\usepackage{color}

\usepackage{perpage}

\MakePerPage{footnote}

\begin{document}
\title[]{Inequivalence of nonequilibrium path ensembles: the example of stochastic bridges}

\author{J Szavits-Nossan$^1$, M R Evans$^1$}

\address{$^1$ SUPA, School of Physics and Astronomy, University of Edinburgh, Peter Guthrie Tait Road, Edinburgh EH9 3FD, United Kingdom}

\ead{jszavits@staffmail.ed.ac.uk}
\ead{mevans@staffmail.ed.ac.uk}

\begin{abstract}
We study stochastic processes in which the trajectories are constrained so that the process realises a large deviation of the unconstrained process. In particular we consider stochastic bridges and the question of inequivalence of path ensembles between the microcanonical ensemble, in which the end points of the trajectory are constrained, and the canonical or s ensemble in which a bias or tilt is introduced into the process. We show how ensemble inequivalence can be manifested by the phenomenon of temporal condensation in which the large deviation is realised in a vanishing fraction of the duration
(for long durations). For diffusion processes we find that  condensation happens whenever the process is subject to a confining potential, such as for the Ornstein-Uhlenbeck process, but not in the borderline case of dry friction in which there is partial ensemble equivalence. We also discuss continuous-space, discrete-time random walks for which in the case of a heavy tailed step-size distribution it is known that the large deviation may be achieved in a single step of the walk. Finally we consider possible effects of several constraints on the process and in particular give an alternative explanation of the interaction-driven condensation in terms of constrained Brownian excursions. 
\end{abstract}






\submitto{\JSTAT}

\date{19/08/2015}

\maketitle

\section{Introduction}
\label{intro}

The problem of extending the thermodynamic formalism to systems out of equilibrium is a central challenge in statistical physics. Significant progress on this topic has been made in the past two decades, yielding several remarkably general results such as fluctuation theorems \cite{EvansCohenMorriss93,EvansSearles94,GallavottiCohen95,Jarzynski97,Kurchan98}. This has resulted in a large body of work that extends the notion of statistical ensembles to dynamical trajectories \cite{Maes99,Crooks00,Seifert05,Lecomte07,HarrisSchutz07,JackSollich10,ChetriteTouchette13,ChetriteTouchette14}, and is closely connected to the mathematical theory of large deviations, which deals with probabilities of rare events (for a review, see e.g. \cite{Touchette09}). In particular, one is interested in calculating the statistics of time-integrated observables $A_T[x]$, where $x$ is a (phase-space) trajectory of duration $T$,
\begin{equation}
P(A_T=a)=\int\mathcal{D}[x]P[x]\delta(A_T[x]-a).
\end{equation}
\noindent Here $P[x]$ is a probability density functional of a path $x$, and the integration is over all possible paths.
%
%
%
Examples of $A_T$ include the action functional \cite{LebowitzSpohn99},
average drift of a Brownian particle \cite{Evans04}, transition history \cite{Lecomte07} and time-integrated particle current in driven diffusive systems, e.g. in exclusion processes \cite{BodineauDerrida04}.
%

Often, it is of interest not only to calculate the probability density $P(A_T=a)$ of a rare event, but to understand how the particular fluctuation leading to the rare event occurred. This amounts to selecting trajectories $x$ that have fixed $A_T[x]=a$, 
\begin{equation}
\label{microcanonical}
P[x\vert A_T=a]=\frac{P[x,A_T=a]}{P(A_T=a)},
\end{equation}
\noindent which resembles the microcanonical ensemble. Unless we are in the low-noise regime, there will be many trajectories contributing to the event $A_T=a$. Selecting trajectories leading to $A_T=a$ from the original, unconstrained dynamics may prove difficult if $A_T=a$ is a rare event (a large deviation), in which case the unconstrained average $\langle A_T\rangle$ is different from $a$ for large $T$, where $\langle\dots\rangle$ is taken with respect to $P[x]$. Even worse, we do not generally expect to find the constrained process to be Markovian, which can pose serious difficulties for its analysis. A resolution to this problem in the spirit of equilibrium statistical physics, that is attracting much interest of late, is to look at the canonical path ensemble instead,
\begin{equation}
\label{canonical}
P_{s}[x]=\frac{P[x]\rme^{sA_T[x]}}{\left\langle \rme^{sA_T}\right\rangle}.
\end{equation}
\noindent This path ensemble is known under several names, depending on the field of study. In physics, it has been called {\it s ensemble} (related to the choice of letter s for the tilting parameter), but also {\it driven} or {\it biased} ensemble; in rare-event simulations it is called {\it (exponentially) tilted} ensemble, or {\it Esscher transform of $P$}. It has been used to probe rare trajectories in systems with metastable states, namely to study the glass transition \cite{MerolleChandler05,JackGarrahanChandler06,GarrahanWijland07,HedgesChandler09,GarrahanWijland09,ChandlerGarrahan10,JackGarrahan10,SpeckGarrahan11} and more recently to address the problem of protein folding \cite{WeberPande13,MeyGarrahan14,WeberPande14}.

Recently, Chetrite and Touchette have rigorously proved several remarkable properties of the canonical path ensemble \cite{ChetriteTouchette13,ChetriteTouchette14}. First, the canonical path ensemble can be realized by a Markov process (referred to as driven dynamics) in the long-time limit. Second, the microcanonical and the canonical path ensembles are asymptotically equivalent in the limit $T\rightarrow\infty$ in the sense that 
\begin{equation}
\label{equivalence}
\lim_{T\rightarrow\infty}\frac{1}{T}\log\frac{P[x\vert A_T=a]}{P_s[x]}=0\;,
\end{equation}
where the limit is approached almost surely and $P[x\vert A_T=a]$ and $P_s[x]$ are given by (\ref{microcanonical}) and (\ref{canonical}) respectively. This important result allows one to study (generally unknown) dynamics of constrained systems by studying dynamics of the driven process, which can be then analyzed using standard Monte Carlo techniques. The equivalence holds under the following three conditions \cite{ChetriteTouchette14,Touchette15}:
\begin{itemize}
\item {\bf Condition A}: $P(A_T=a)$ must satisfy a large deviation principle with a rate function $I(a)$,
\begin{equation}
\label{ld}
P(A_T=a)\asymp \rme^{-TI(a)}, \quad T\rightarrow\infty,
\end{equation}
\noindent where $\asymp$ denotes asymptotic behaviour in the sense that
\begin{equation}
I(a)=-\lim_{T\rightarrow\infty}\case{1}{T}\log P(A_T=a),
\label{Idef}
\end{equation}
\item {\bf Condition B}: the large deviation $A_T=a$ must arise entirely from the interior of the interval $[0,T]$, e.g. excluding large deviations at the boundaries $0$ or $T$,
\item {\bf Condition C}: the rate function $I(a)$ must be convex at $a$.
\end{itemize}
\noindent We note that the rate function $I(a)$ does not need to be differentiable at $a$ for the equivalence to hold. However, if it is then the tilting parameter is unique and is given by $s=I'(a)$. 
Conditions {\bf A, B, C} are found to be sufficient  for
the ensemble equivalence to hold; whether all conditions are necessary 
remains to be clarified.

In this work, we are interested in cases where the microcanonical and canonical path ensembles are not equivalent. The inequivalence of the nonequilibrium path ensembles has been recently reported for large current fluctations in the zero-range process \cite{HarrisRakosSchutz06,RakosHarris08}. The exactly solvable one-site model revealed that the problem lies in the infinite state space.
For example, the spectrum of the driven process may become gapless or
the process may  have non-normalizable eigenvectors with respect to the initial measure.  Beyond this example, the necessary 
properties of the  spectrum required for the equivalence of nonequilibrium ensembles is an open problem. 

In equilibrium, the inequivalence of microcanonical and canonical ensembles has been extensively studied for various models (for a review, see \cite{CampaRuffo09}), typically for systems with long-range interactions, such as gravitational systems, dipolar or unscreened Coulomb systems. The total energy in these systems is no longer extensive in the energies of the subsystems, which may lead to the microcanonical entropy that is a nonconcave function of the energy. In that case the canonical free energy is not differentiable at some point and thus there is no temperature to establish the connection with the canonical ensemble \cite{Touchette11}. Consequently, these systems show interesting phenomena such as negative (microcanonical) specific heat, slow relaxation, quasi-steady states, etc.

This motivates us to pose the following questions. What are the typical situations, analogous to e.g. long-range interactions in equilibrium, where the equivalence of path ensembles is not expected to hold? Moreover, are there common phenomena arising from the inequivalence of nonequilibrium path ensembles? 

To address these questions we focus here on stochastic bridges \cite{Doob57,Orland11,MajumdarOrland15}, which are obtained by conditioning Markov processes on fixed $x_0$ and $x_T$. 
Thus the observable $A_T$ is given by
\begin{equation}
A_T=\frac{x(T)-x(0)}{T}\;.
\label{AT1}
\end{equation}
Crucially, stochastic bridges are themselves Markov processes for any finite time $T$ (a property that is unlikely to hold for conditioning on more general $A_T$) which allows us to study their dynamics analytically. We present the solutions of several stochastic bridges for which the original, unconstrained dynamics violates one of the conditions A or B for ensemble equivalence mentioned above. These exact solutions shed light on why ensemble equivalence breaks down. For the examples we consider, ensemble inequivalence is typically manifested by condensation-like phenomena: the conditioning $A_T=a$ causes the original process to `condense' its large deviation (to meet the conditioning) in a vanishing fraction of the interval $[0,T]$, rather than throughout the whole interval as implied by the driven process associated to the canonical ensemble (\ref{canonical}).

Related condensation phenomena are well known in the problem of sums of independent and identically distributed random variables. If the distribution of the random variables is heavy tailed then a large deviation of the sum typically occurs through a single random variable realising the  large deviation of the sum \cite{Linnik61,Nagaev69,GrosskinskySchutzSpohn03,EMZ06}. Moreover, recent work has shown that a heavy-tailed distribution is not necessary for condensation to occur when there is a further constraint on the random variables in addition to their sum \cite{SNEM14a,SNEM14b}. This type of condensation is manifested in real-space condensation in spatially extended stochastic mass transport models such as the zero-range process wherein a single site captures a finite fraction of the total mass in the system \cite{EvansHanney05}. In this work we will examine the connection between temporal condensation exhibited in discrete-time stochastic bridges and condensation within collections of discrete random variables.

We start by reviewing known results for diffusion bridges previously presented in \cite{ChetriteTouchette14}: the equivalence of ensembles holds for Brownian motion, but fails for the Ornstein-Uhlenbeck process where the large deviation is condensed into a boundary effect. We extend this result for temporal condensation to other diffusion processes in which there is a form of potential that causes the conditioning $x_T=aT$ to be met in a non-extensive manner, to be explained in detail later. To stress the importance of the condition B for the equivalence to hold, we present a diffusion bridge for
a multiplicative noise process (the CIR bridge)  in  which $P(A_T=a)$ has an exponential tail, but where the large deviation $A_T=a$ is again a boundary effect. As a borderline case, we also study the Brownian motion with dry friction \cite{deGennes05,TouchetteJust10}, for which the path ensembles are equivalent, but the connection between the two ensembles is not unique - a case known as partial equivalence \cite{EllisHavenTurkington00,Touchette11}. 

After studying diffusion bridges, we look at several discrete-time Markov chains driven by heavy-tailed (and thus non-Gaussian) noise. There the rate function is formally zero and thus violates the condition A, which can be also related to the phenomenon of condensation in stationary states of mass-transport models (for a review, see \cite{EvansHanney05}). At the end, we use the equivalence (\ref{equivalence}) to revisit the interaction-driven condensation \cite{EvansHanneyMajumdar06}, and show that it is equivalent to the conditioning of random walk trajectories on a large deviation of their local time, which counts the number of returns to the origin. Our results should serve as guiding principle for other, more complex stochastic systems, for which conditions A or B are generally difficult to examine.

The paper is organised as follows. We study stochastic bridges for diffusion processes in Section \ref{diffusion_bridges} and for discrete-time Markov chains driven by heavy-tailed noise in Section \ref{rw_bridges}. Discussion and conclusions based on these examples on when not to expect the equivalence to hold are presented in Section \ref{conclusions}.


\section{Diffusion bridges}
\label{diffusion_bridges}

In this paper, we consider a diffusion process defined by the following (It\^{o}) stochastic differential equation (SDE),
\begin{equation}
\label{sde}
\frac{dx}{dt}=b(x,t)+\sigma(x,t)\eta(x,t), \quad x(0)=0
\end{equation}
\noindent where $\eta(t)$ is $\delta$-correlated noise, 
\begin{equation}
\label{noise}
\langle\eta(t)\rangle=0, \quad \langle\eta(t)\eta(t')\rangle=\delta(t-t').
\end{equation}
\noindent A diffusion bridge is a process obtained by conditioning $x(t)$ in (\ref{sde}) to have a fixed value of $K$ at time $T$\footnote{This value is traditionally set to $0$; here we adopt the name ``bridge'' for a diffusion process on $[0,T]$ conditioned on $x(T)=K$, irrespective of the value of $K$.}. It is an ideal candidate for studying nonequilibrium path ensembles, for two reasons. First, the  observable  $A_T$ (\ref{AT1}) takes the simple form 
\begin{equation}
A_T =\frac{K}{T}=\frac{aT}{T}=a,
\label{AT2}
\end{equation}
where we have set $K=aT$ to impose a large deviation. Second, the conditioned process itself can be conveniently described by a stochastic differential equation that is similar to (\ref{sde}), but with a modified drift term \cite{Doob57,Orland11,ChetriteTouchette14,MajumdarOrland15}. To see this, let $p(x,t\vert 0,0)$ solve the Fokker-Planck equation corresponding to the stochastic differential equation (\ref{sde}), subject to the initial condition $p(x,0\vert 0,0)=\delta(x)$,
\begin{equation}
\label{fp}
\frac{\partial p(x,t|0,0)}{\partial t}=-\frac{\partial}{\partial x}\left[b(x,t)p\right]+\frac{1}{2}\frac{\partial^2}{\partial x^2}\left[\sigma^2(x,t) p\right].
\end{equation}
\noindent The probability density $g(x,t)$ for the stochastic bridge is given by
\begin{equation}
g(x,t)=\frac{p(K,T\vert x,t)p(x,t\vert0,0)}{p(K,T\vert 0,0)}.
\end{equation}
\noindent Here, $p(K,T\vert x,t)$ solves the backward Kolmogorov equation
\begin{equation}
\label{backward}
\frac{\partial }{\partial t}p(K,T\vert x,t)=-b(x,t)\frac{\partial p}{\partial x}-\frac{1}{2}\sigma^2(x,t)\frac{\partial^2 p}{\partial x^2},
\end{equation}
\noindent subject to the final condition $p(K,T\vert x,T)=\delta(K-x)$. 

Now  it is straightforward to obtain
\begin{equation}
\frac{\partial g(x,t)}{\partial t}
=\frac{p(K,T\vert x,t)}{p(K,T\vert 0,0)} \frac{\partial p(x,t\vert0,0)}{\partial t} +
\frac{p(x,t\vert0,0)}{p(K,T\vert 0,0)} \frac{\partial p(K,T\vert x,t) }{\partial t}.
\end{equation}
Then substituting (\ref{fp},\ref{backward}) and regrouping terms yields, after some calculation, the following Fokker-Planck equation for $g(x,t)$
\begin{equation}
\label{fp_bridge}
\frac{\partial g(x,t)}{\partial t}=-\frac{\partial}{\partial x}\left[c_T(x,t)g\right]+\frac{1}{2}\frac{\partial^2}{\partial x^2}\left[\sigma^2(x,t) g\right],
\end{equation}
where $c_T(x,t)$ is given by
\begin{equation}
\label{drift_bridge}
c_T(x,t)=b(x,t)+\sigma^2(x,t)\frac{\partial}{\partial x}\ln p(K,T\vert x,t),
\end{equation}
\noindent and the subscript $T$ emphasises that $c_T(x,t)$ depends explicitly on $T$. Using (\ref{fp_bridge}), we arrive at the following stochastic differential equation for the diffusion bridge which we denote here and in the following by $y(t)$
\begin{equation}
\label{sde_bridge}
\frac{dy}{dt}=c_T(y,t)+\sigma(y,t)\eta(y,t),
\end{equation}
\noindent where the expression for $c_T(x,t)$ is given in (\ref{drift_bridge}). Note that the equation (\ref{sde_bridge}) is an exact equation for the conditioned process which is a diffusion with (non-homogeneous) drift given by (\ref{drift_bridge}) and noise width  $\sigma$. This fact allows us to study the  microcanonical dynamics, provided we can solve the backward Kolmogorov equation for $p(K,T;x,t)$. If so, the equation (\ref{sde_bridge}) can be then solved numerically using e.g. an Euler-Mayurama scheme. Moreover, if  $p(K,T;x,t)$ has a simple enough form we can  compute exactly  various  quantities  for the conditioned process. Next, we will present several examples of diffusion bridges, for which there is an explicit expression for $g(x,t)$ and thus for $c_T(x,t)$.


\subsection{Brownian bridge}
\label{brownian_bridge}

For the Brownian bridge discussed e.g. in Ref. \cite{ChetriteTouchette14}, we have 
\begin{equation}
b(x,t)=\textrm{const.}\equiv \mu,\quad \sigma(x,t)=\textrm{const.}\equiv\sigma,
\end{equation}
\noindent which also includes a special case of $\mu=0$ called the Wiener bridge. The solution to the corresponding Fokker-Planck equation subject to $p(x,0)=\delta(x)$ is given by
\begin{equation}
\label{fp_solution_brownian}
p(x,t\vert 0,0)=\frac{1}{\sigma\sqrt{2\pi t}}\rme^{-\case{(x-\mu t)^2}{2 t\sigma^2}}.
\end{equation}
\noindent Because the process is both space- and time-homogeneous, the expression for $p(K,T\vert x,t)$ is the same as for $p(K-x,T-t\vert 0,0)$, yielding
\begin{equation}
p(K,T\vert x,t)=\frac{1}{\sigma\sqrt{2\pi (T-t)}}\rme^{-\case{[K-x-\mu (T-t)]^2}{2 (T-t)\sigma^2}}.
\end{equation}
\noindent For $K=at$, the stochastic differential equation for the Brownian bridge is thus given by
\begin{equation}
\label{sde_brownian_bridge}
\frac{dy}{dt}=\frac{aT-y}{T-t}+\sigma\eta,
\end{equation}
where we denoted the bridge by $y(t)$ to distinguish it from the unconstrained process. We note that the drift $(aT-y)/(T-t)$ of the bridge process is actually time and space dependent, but as we shall see, for large $y$ becomes independent of $T$ and $\simeq a$. Also note that equation (\ref{sde_brownian_bridge}) is linear in $y$ and therefore its solution can be written explicitly,
\begin{equation}
\label{brownian_bridge_solution}
y(t)= at+\sigma(T-t)\int_{0}^{t}\frac{\rmd W_s}{T-s},
\end{equation}
\noindent where $W_s$ is Wiener process. The process $y(t)$ in (\ref{brownian_bridge_solution}) is Gaussian, whose mean and variance can be easily evaluated from (\ref{brownian_bridge_solution}) and read
\begin{equation}
\label{mean_var_brownian_bridge}
\langle y(t)\rangle=a t,\quad \langle y(t)^2\rangle-\langle y(t)\rangle^2=\sigma^2 t\left(1-\frac{t}{T}\right).
\end{equation}

%
%
\begin{figure}[hbt]
\centering\includegraphics[width=8cm]{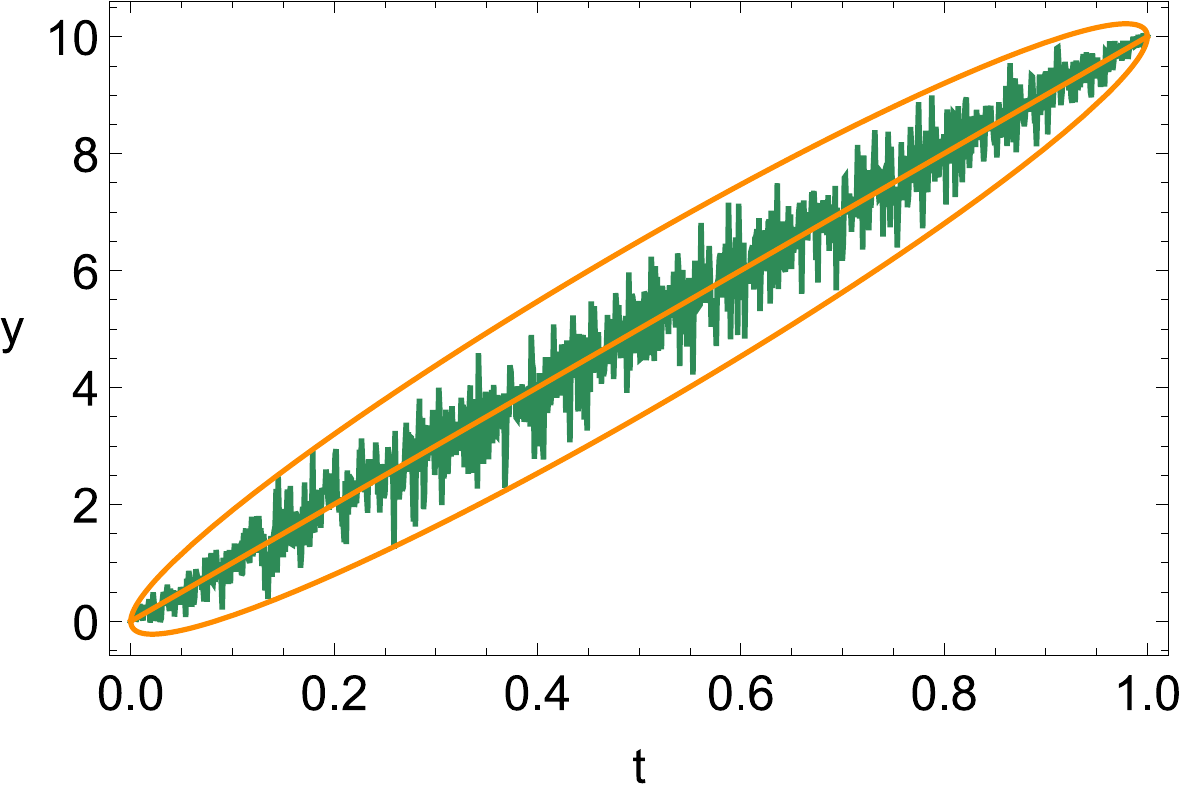}
\caption{Brownian bridge starting at $y(0)=0$ and ending at $y(T)=aT$ for $\mu=0$, $\sigma=1$, $a=10$ and $T=1$, calculated at discrete time intervals of size $\epsilon=0.001$; solid lines are $\langle y\rangle$ and $\langle y\rangle \pm 3[\langle y^2\rangle-\langle y\rangle^2]^{1/2}$, where $\langle y\rangle$ and $\langle y^2\rangle-\langle y\rangle^2$ are given in (\ref{mean_var_brownian_bridge}).} 
\label{fig1}
\end{figure}

\noindent A sample trajectory for the constrained process, obtained at discrete time intervals of size $\epsilon=0.001$, is presented in Figure \ref{fig1}.

The equivalence between the Brownian bridge and the corresponding driven process has been established in Ref. \cite{ChetriteTouchette14}. Essentially, one looks at the probability density $P(A_T=a)=P(x(T)/T=a)$ for the unconstrained dynamics, which from (\ref{fp_solution_brownian}) reads
\begin{equation}
P\left(A_T=\case{x(T)}{T}=a\right)=\frac{1}{\sigma\sqrt{2\pi T}}\rme^{-T\case{(a-\mu)^2}{2\sigma^2}}.
\end{equation}
\noindent The corresponding rate function is given by
\begin{equation}
\label{rate_brownian}
I(a)=\frac{(a-\mu)^2}{2\sigma^2},
\end{equation}
\noindent and is differentiable, yielding $s=I'(a)=\case{a-\mu}{\sigma^2}$ as the tilting parameter. One then constructs `driven dynamics' using the generalised Doob transform \cite{ChetriteTouchette14}, to obtain that the drift of the driven process in the large $T$ limit is exactly $a$, as for the Brownian bridge.

The Brownian bridge is an example where the (global) conditioning modifies trajectories locally in the interior of the interval $[0,T]$. The next example - that of the Ornstein-Uhlenbeck bridge - is rather different, in the sense that the large deviation that fulfils the conditioning is concentrated at the end of the interval.


\subsection{Ornstein-Uhlenbeck bridge}
\label{OU_bridge}

\begin{figure}[hbt]
\centering\includegraphics[width=8cm]{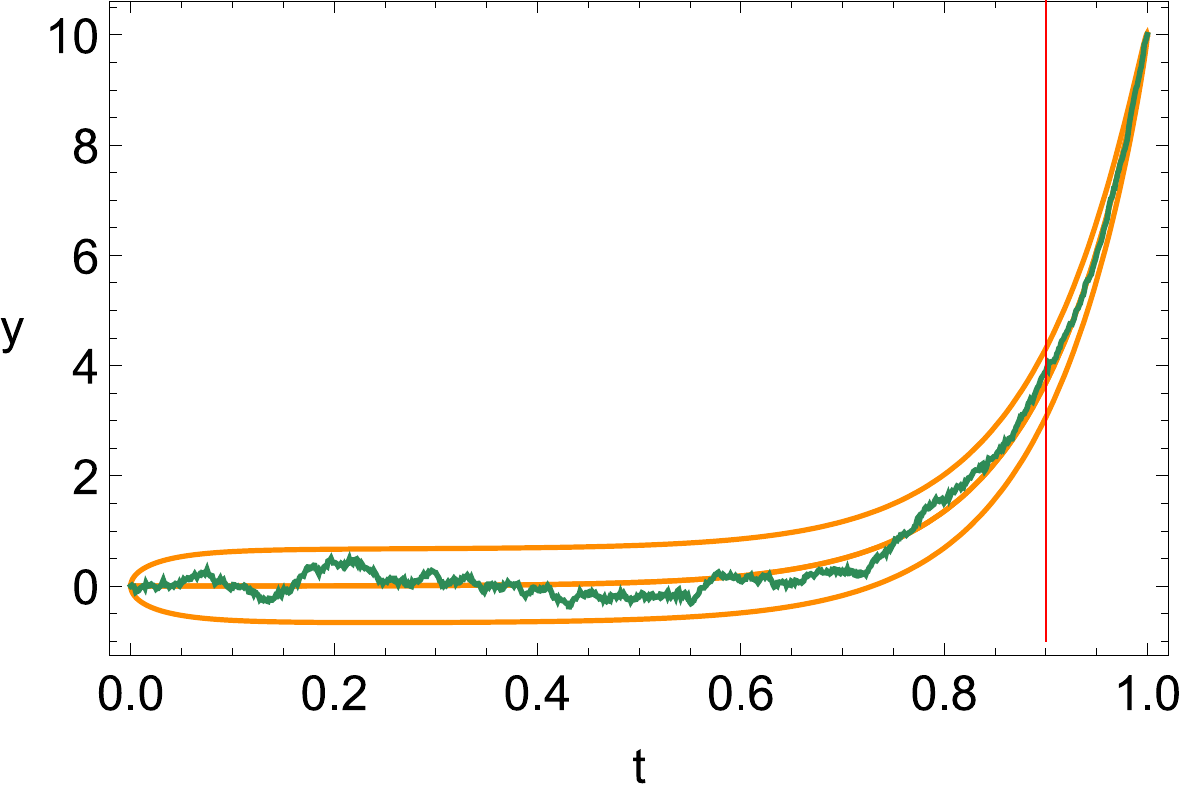}
\caption{A sample trajectory of the Ornstein-Uhlenbeck bridge starting at $y(0)=0$ and ending at $y(T)=aT$ for $\theta=10$, $\mu=0$, $\sigma=1$, $a=10$ and $T=1$, calculated at discrete time intervals of size $\epsilon=0.001$; solid lines are $\langle y\rangle$ and $\langle y\rangle \pm 3[\langle y^2\rangle-\langle y\rangle^2]^{1/2}$, where $\langle y\rangle$ and $\langle y^2\rangle-\langle y\rangle^2$ are given in (\ref{mean_var_ou_bridge}). Vertical line is placed at $T-1/\theta$ and denotes a characteristic time scale over which the large deviation occurs.}
\label{fig2}
\end{figure}

Next we review results for the Ornstein-Uhlenbeck process, also studied in Ref. \cite{ChetriteTouchette14}, for which 
\begin{equation}
b(x,t)=\theta(\mu-x),\quad \sigma(x,t)=\textrm{const.}\equiv\sigma,
\end{equation}
\noindent where $\theta>0$. The solution to the corresponding Fokker-Planck equation subject to $p(x,0)=\delta(x)$ is given by
\begin{equation}
\label{fp_solution_ou}
p(x,t\vert 0,0)=\frac{\rme^{-\case{\left[x-\mu\left(1-\rme^{-\theta t}\right)\right]^2}{2 \sigma^2(1-\rme^{-2\theta t})/(2\theta)}}}{\sigma\sqrt{2\pi(1-\rme^{-2\theta t})/(2\theta)}}.
\end{equation}
\noindent Similarly, the expression for $p(K,T\vert x,t)$ is given by 
\begin{equation}
p(K,T\vert x,t)=\frac{\rme^{-\case{\left[K-\mu- (x-\mu)\rme^{-\theta(T-t)}\right]^2}{2 \sigma^2[1-\rme^{-2\theta (T-t)}]/(2\theta)}}}{\sigma\sqrt{2\pi[1-\rme^{-2\theta (T-t)}]/(2\theta)}}.
\end{equation}
\noindent For $K=at$, the stochastic differential equation for the Ornstein-Uhlenbeck bridge \cite{ChetriteTouchette14}, denoted $y(t)$, is thus given,
from  (\ref{drift_bridge}) and (\ref{sde_bridge}), by
\begin{equation}
\label{sda_ou_bridge}
\frac{dy}{dt}=\theta(\mu-y)\frac{\cosh(\theta(T-t))}{\sinh(\theta(T-t))}+\frac{\theta(aT-\mu)}{\sinh(\theta(T-t))}+\sigma\eta.
\end{equation}
\noindent This equation is again linear in $y$ and its solution can be written explicitly as
\begin{eqnarray}
y(t)&=&(aT-\mu)\frac{\sinh(\theta t)}{\sinh (\theta T)}+\mu\left[1-\frac{\sinh(\theta(T-t))}{\sinh(\theta T)}\right]+\nonumber\\
&& +\sigma\sinh(\theta (T-t))\int_{0}^{t}\frac{\rmd W_s}{\sinh(\theta(T-s))}.
\end{eqnarray}
\noindent As a Gaussian process, $y(t)$ is fully determined by its mean and variance which read
\begin{eqnarray}
\label{mean_var_ou_bridge}
&&\fl\qquad \langle y(t)\rangle=(aT-\mu)\frac{\sinh(\theta t)}{\sinh (\theta T)}+\mu\left[1-\frac{\sinh(\theta(T-t))}{\sinh(\theta T)}\right],\\
&&\fl\qquad \langle y(t)^2\rangle-\langle y(t)\rangle^2=\sigma^2\frac{\sinh(\theta (T-t))\sinh(\theta t)}{\sinh(\theta T)}.
\end{eqnarray}
\noindent A sample trajectory for the constrained process, obtained at discrete time intervals of size $\epsilon$ = 0.001, is presented in Figure \ref{fig2}.

%
%

For $\theta t,\theta T\gg 1$, the average of $y(t)$ can be approximated by a simpler expression
\begin{equation}
\langle y(t)\rangle \approx (aT-\mu)\rme^{-\theta(T-t)}+\mu\left(1-\rme^{-\theta t}\right), \quad \theta t,\theta T\gg 1.
\end{equation}
\noindent In contrast to the Brownian bridge $y(t)$ in Figure \ref{fig1} where $\langle y(t)\rangle=at$, here $\langle y(t)\rangle$ has the same value $\approx \mu$ as in the unconstrained process, except for $t$ approximately $1/\theta$ away from $0$ and $T$. Put differently, the conditioning to reach $aT$ at time $T$ has no effect on the stochastic dynamics in the interior of $[0,T]$, except for a small fraction of time close to $T$, which goes to $0$ as $T\rightarrow\infty$. This boundary effect is not recovered by the driven process, which has the same drift as the unconstrained process \cite{ChetriteTouchette14}. The driven process correctly describes the conditioned process in the interior of $[0,T]$, but not at the right boundary.

Here we give an intuitive explanation for this effect. The time-integrated observable $A_T$, such as $A_T=[x(T)-x(0)]/T$, is nothing more than a sum (or an integral) of random variables. Loosely speaking, the large deviation $A_T=a$ is an interior effect if all of these random variables can be changed locally to achieve the large deviation. To see that this is not the case in the Ornstein-Uhlenbeck process, we consider its discrete-time Markov chain analogue, called the autoregressive process of order $1$, or AR(1),
\begin{equation}
\label{AR1_eq}
X_t=X_{t-1}+\theta(\mu-X_{t-1})+\sigma\eta_t,\quad X_0=0
\end{equation}
\noindent where we assume that $0<\theta<1$, and $\eta_t$ are independent and identically distributed Gaussian random variables with the mean $\langle\eta\rangle=0$ and variance $\langle\eta^2\rangle=1$. The stochastic recurrence equation (\ref{AR1_eq}) can be iterated yielding
\begin{equation}
\label{X_T_AR1}
X_T=\mu[1-(1-\theta)^T]+\sigma\sum_{t'=1}^{T}(1-\theta)^{T-t'}\eta_{t'}.
\end{equation}

\noindent In principle, all $\eta_{t'}$ contribute to a large deviation of their sum. However, the weighting factor $(1-\theta)^{T-t'}$ will make only few of them contribute to the sum, ones that are close to $T$. In this sense, we may say that the boundary effect is due to the fact that $X_T$ is not extensive in $T$ - increasing $T$ will not increase the number of random variables $\eta_{t'}$ that contribute substantially to $X_T$. From (\ref{fp_solution_ou}) we also get for $P(x(T)/T=a)$ the following expression,
\begin{equation}
P\left(\case{X_T}{T}=a\right)\propto\rme^{-\case{\theta a^2 T^2}{\sigma^2}}, \quad T\rightarrow\infty,
\end{equation}
\noindent so that the rate function $I(a)$ defined with respect to the limit $T\rightarrow\infty$ in (\ref{Idef}) is formally infinite. However, it is important to emphasise that the main reason the path equivalence (\ref{equivalence}) does not hold here is really the boundary effect described above. In the next example, we will study a process for which $P(x(T)/T=a)$ has an exponential tail, but the equivalence does not hold because of a similar boundary effect.


\subsection{Cox-Ingersoll-Ross (CIR) bridge}
\label{CIR_bridge}
Here we study a process that has the same drift as the Ornstein-Uhlenbeck process, but with a different, state-dependent diffusion coefficient

\begin{equation}
b(x,t)=\theta(\mu-x),\quad \sigma(x,t)=\sigma\sqrt{x}, \quad x\geq 0.
\end{equation}

\begin{figure}[hbt]
\centering\includegraphics[width=8cm]{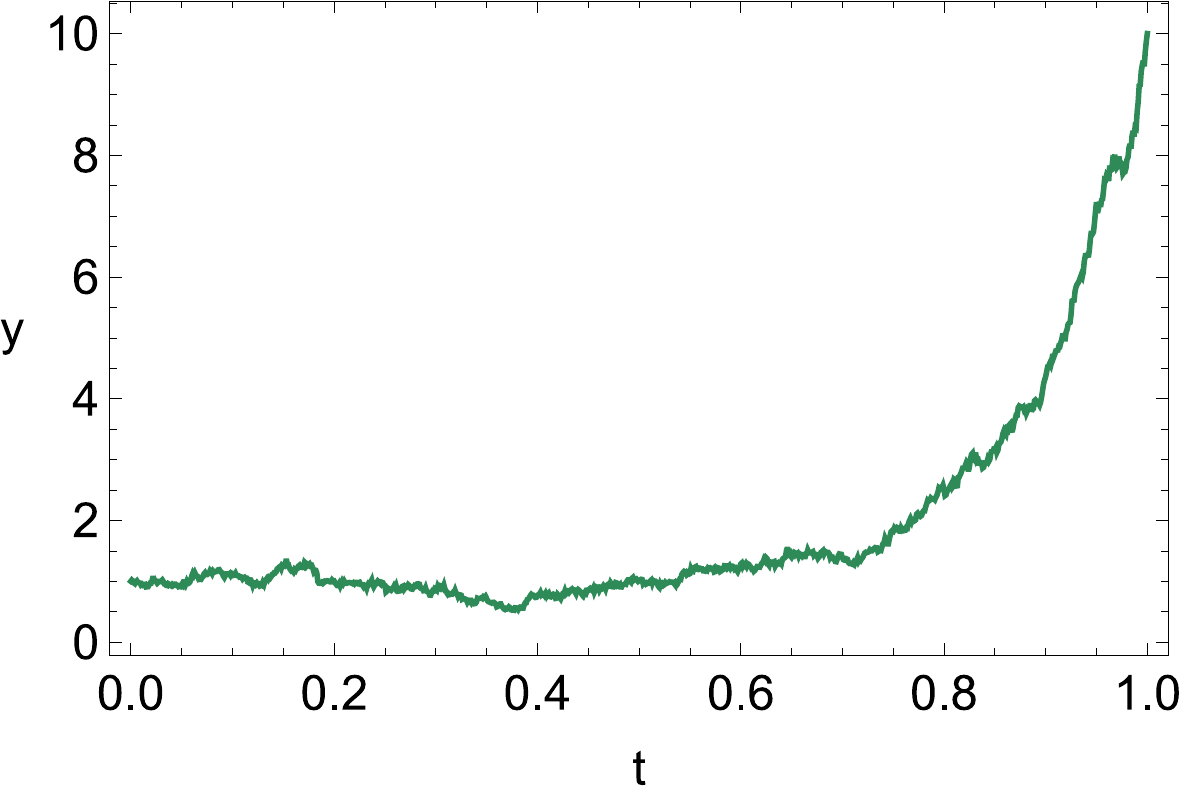}
\caption{A sample trajectory of the Cox-Ingersoll-Ross bridge starting at $y(0)=1$ and ending at $y(T)=aT$ for $\theta=10$, $\mu=1$, $\sigma=1$, $a=10$ and $T=1$, calculated at discrete time intervals of size $\epsilon=0.001$.}
\label{fig3}
\end{figure}

\noindent This process is known as the Cox-Ingersoll-Ross model or the CIR process, and belongs to a class of square-root diffusions. The CIR process is a popular model for the evolution of interest rates in finance \cite{CIR}.
Remarkably, the propagator $p(x,t\vert x',t')$ of the CIR process is known in a closed form \cite{Feller},
\begin{eqnarray}
\label{CIR_propagator}
p(x,t\vert x',t')&=&\frac{2\theta}{\sigma^2[1-e^{-\theta(t-t')}]}\left(\frac{x}{x'e^{-\theta(t-t')}}\right)^{q/2}e^{-\frac{2\theta(x+x'e^{-\theta(t-t')})}{\sigma^2[1-e^{-\theta(t-t')}]}}\nonumber\\
&&\times I_q\left(\frac{4\theta\sqrt{x x' e^{-\theta(t-t')}}}{\sigma^2[1-e^{-\theta(t-t')}]}\right),
\end{eqnarray}
\noindent where $q=(2\theta\mu/\sigma^2-1)$ and $I_q(x)$ is modified Bessel function of the first kind. The expression for $p(aT,T\vert x,t)$ can be then used to calculate the drift $c_T(x,t)$, and the corresponding SDE for the bridge can be then integrated numerically. However, we note that since $x$ must be non-negative at all times, the standard Euler-Mayurama scheme is not appropriate for numerical integration. Here, we used modified Euler scheme for SDEs with square-root diffusion coefficient described in Ref. \cite{BerkaouiBossyDiop2008}, which has strong convergence. A sample trajectory for the constrained process, obtained at discrete time intervals of size $\epsilon$ = 0.001, is presented in Figure \ref{fig3}. Sample trajectories of the CIR process look similar to those for the Ornstein-Uhlenbeck process, the difference being that larger values of $x$ are more stochastic, which is due to the factor $\sqrt{x}$ multiplying white noise term. On the other hand, the limit
\begin{equation}
I(a)=\lim_{T\rightarrow\infty}\frac{1}{T}\textrm{ln}P\left(\frac{x(T)}{T}=a\right)=\frac{2\theta a}{\sigma^2}
\end{equation}
is finite, which emphasises the importance of the condition B for establishing the equivalence even when the condition A is satisfied.
%
%

The examples of sections \ref{OU_bridge} and \ref{CIR_bridge} illustrate the phenomenon of `temporal condensation' (the name  has been suggested in \cite{ChetriteTouchette14}), whereby a large deviation is localised to a small fraction of the time interval $T$ (that vanishes in the limit $T\rightarrow\infty$), rather than distributed over the whole time interval as for the Brownian motion in Section \ref{brownian_bridge}.

Next, we consider a borderline case -- a Brownian motion with dry friction -- which interpolates exactly between the Brownian bridge and the last two cases, yielding a finite fraction of the time interval $T$ that is spent around the mean value of the unconstrained process.


\subsection{Brownian bridge with dry friction}
\label{dry_bridge}

Brownian motion with dry (or Coulomb) friction is governed by the following SDE:
\begin{equation}
\label{browniandry}
m\frac{dv}{dt}=-\gamma v -\Delta_{F}\textrm{sgn}(v)+m\sigma\eta(t),
\end{equation}
where $\gamma$ is a friction coefficient and $\Delta_F$ is a threshold force. Here we consider only the inviscid case $\gamma=0$, which was first proposed by de Gennes to describe a particle on a vibrating plate \cite{deGennes05}; for $\gamma\neq 0$, see work by Touchette, Van der Straeten and Just \cite{TouchetteJust10}. 

By introducing  $\Delta=\Delta_F/m$ and renaming $x=v$, we can write (\ref{browniandry}) with $\gamma=0$ as
\begin{equation}
\label{inviscid_dry}
\frac{dx}{dt}=-\Delta\textrm{sgn}(x)+\sigma \eta,
\end{equation}
which identifies $b(x,t)=-\Delta\textrm{sgn}(x)$ and $\sigma(x,t)=\textrm{const.}=\sigma$. The solution $p(x,t\vert x',0)$ to the corresponding Fokker-Planck equation is known \cite{TouchetteJust10} and is given by
\begin{eqnarray}
\label{fp_solution_dry}
&& p(x,t\vert x',0)=\frac{e^{-\Delta^2t/(2\sigma^2)}}{\sqrt{2\pi\sigma^2t}}e^{-\Delta(\vert x\vert-\vert x'\vert)/\sigma^2}e^{-\frac{(x-x')^2}{2\sigma^2t}}\nonumber\\
&&+\Delta\frac{e^{-2\Delta\vert x\vert/\sigma^2}}{2\sigma^2}\left[1+\textrm{erf}\left(\frac{\Delta t-(\vert x\vert+\vert x'\vert)}{\sqrt{2\sigma^2t}}\right)\right],
\end{eqnarray}
\noindent where $\textrm{erf}(x)$ is the error function. Since the process is time-homogeneous, we can write $p(x,t\vert x',t')=p(x,t-t'\vert x',0)$, yielding the following expression for the drift $c_T(x,t)$ from (\ref{drift_bridge}),
\begin{equation}
\label{drift_bridge_dry}
c_T(x,t)=-\Delta\textrm{sgn}(x)+\frac{\Delta\textrm{sgn}(x)+\case{aT-x}{T-t}-\Delta\textrm{sgn}(x)\rme^{-\case{aT(x+\vert x\vert)}{\sigma^2(T-t)}}}{1+\Delta\sqrt\case{\pi(T-t)}{2\sigma^2}\rme^{-\case{aT(x+\vert x\vert)}{\sigma^2(T-t)}}\rme^{J_{T}^{2}}\textrm{erfc}(J_T)},
\end{equation}
\noindent where $\textrm{erfc}(x)=1-\textrm{erf}(x)$ and $J_T$ is given by
\begin{equation}
\label{I}
J_T(x,t)=\frac{(a-\Delta)T+\vert x\vert+\Delta t}{\sqrt{2\sigma^2 (T-t)}}.
\end{equation}
\noindent We can easily check that for $\Delta=0$ the drift term $c_T(x,t)$ reduces to $(aT-x)/(T-t)$ which is the case of the Brownian bridge (\ref{sde_brownian_bridge}).

First let  us  consider $c_T(x,t)$ for  small $t$  in the limit when $T\rightarrow\infty$ for $x\approx 0$ , in which case $J_T\rightarrow -\infty$ for $a<\Delta$ and $J_T\rightarrow\infty$ for $a>\Delta$.  Using the asymptotic expansion of the complementary error function, $\textrm{erfc}(x)\simeq [\textrm{sgn}(x)-1]+\textrm{exp}(-x^2)/(\sqrt{\pi}x)$ for large $\vert x\vert$, we obtain 
\begin{equation}
c_T(x,t)=\cases{-\Delta\textrm{sgn}(x), & $a<\Delta$ \cr \frac{aT-x}{T-t}, & $a>\Delta$ \cr}.
\end{equation}
\noindent This behaviour of the drift $c_T(x,t)$ implies that, in the beginning, the process starting at $x(0)=0$ behaves in the case $a <\Delta$ as the original, unconstrained process which stays close to $x=0$, or
in the case $a >\Delta$  as a Brownian bridge which on average ascends with drift $a$. 

Let us now consider the case  $a <\Delta$ in which case the process begins like the unconstrained dry friction process. Consider  the  limit when both $T, t  \rightarrow\infty$  with $\tau = t/T$ for $x\approx 0$. Then $J_T\rightarrow -\infty$ for $\tau < 1-a/\Delta$ and $J_T\rightarrow\infty$ for  $\tau > 1-a/\Delta$.  As before, we obtain asymptotically 
\begin{equation}
c_T(x,t)=\cases{-\Delta\textrm{sgn}(x), & 
$\tau < 1-a/\Delta$  \cr \frac{aT-x}{T-t}, & $\tau >1-a/\Delta$\cr}.
\end{equation}
\noindent This behaviour implies that for a fraction of the duration $\tau = 1-a/\Delta$ the process behaves as the original, unconstrained process  which stays close to $x=0$. Then for later times $t > (1-a/\Delta)T$ the process behaves as a Brownian bridge which ascends to the end point $x=aT$. Now in this regime  we have $x \simeq (t-\tau T)c_T$ which yields $x \simeq \Delta(t-\tau T)$ and $c_T \simeq \Delta$. Thus in the Brownian bridge section of the trajectory the average drift is $\Delta$. Note this drift is larger than $a$. 

\begin{figure}[hbt]
\includegraphics[height=4cm]{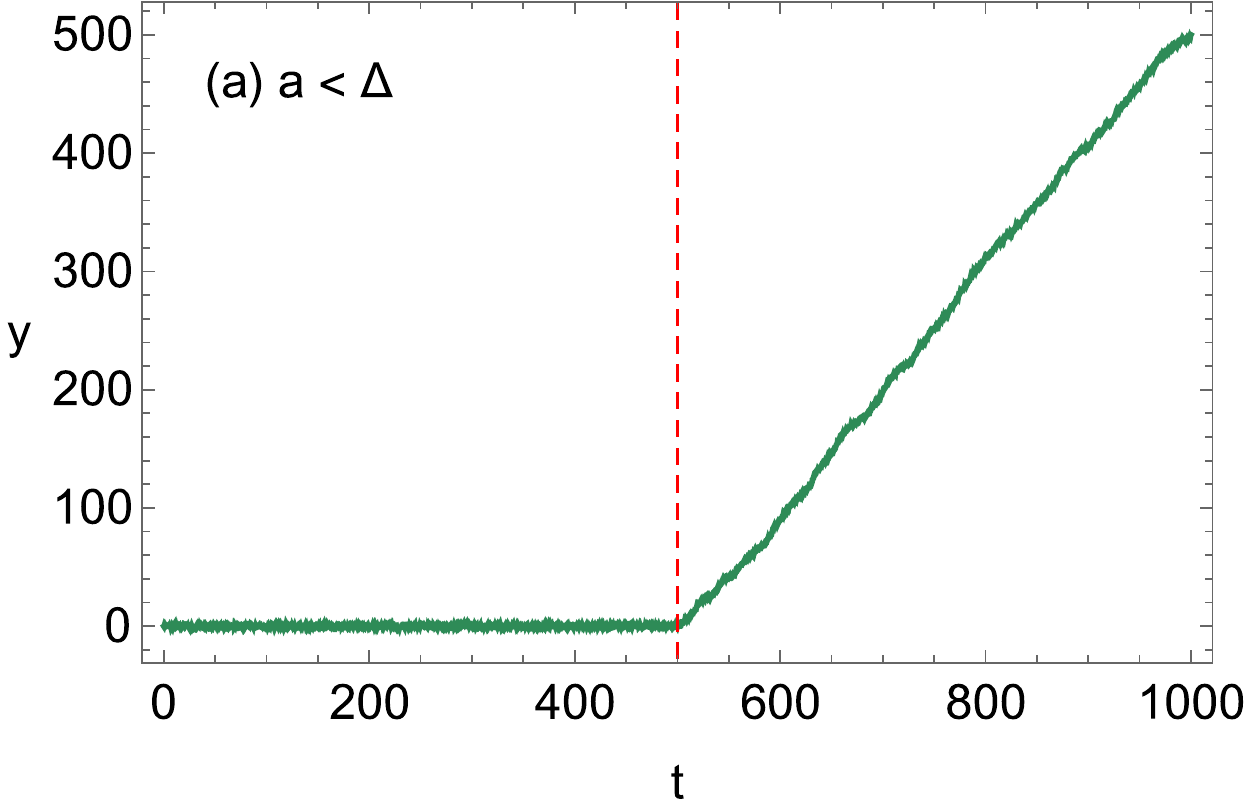}
\includegraphics[height=4cm]{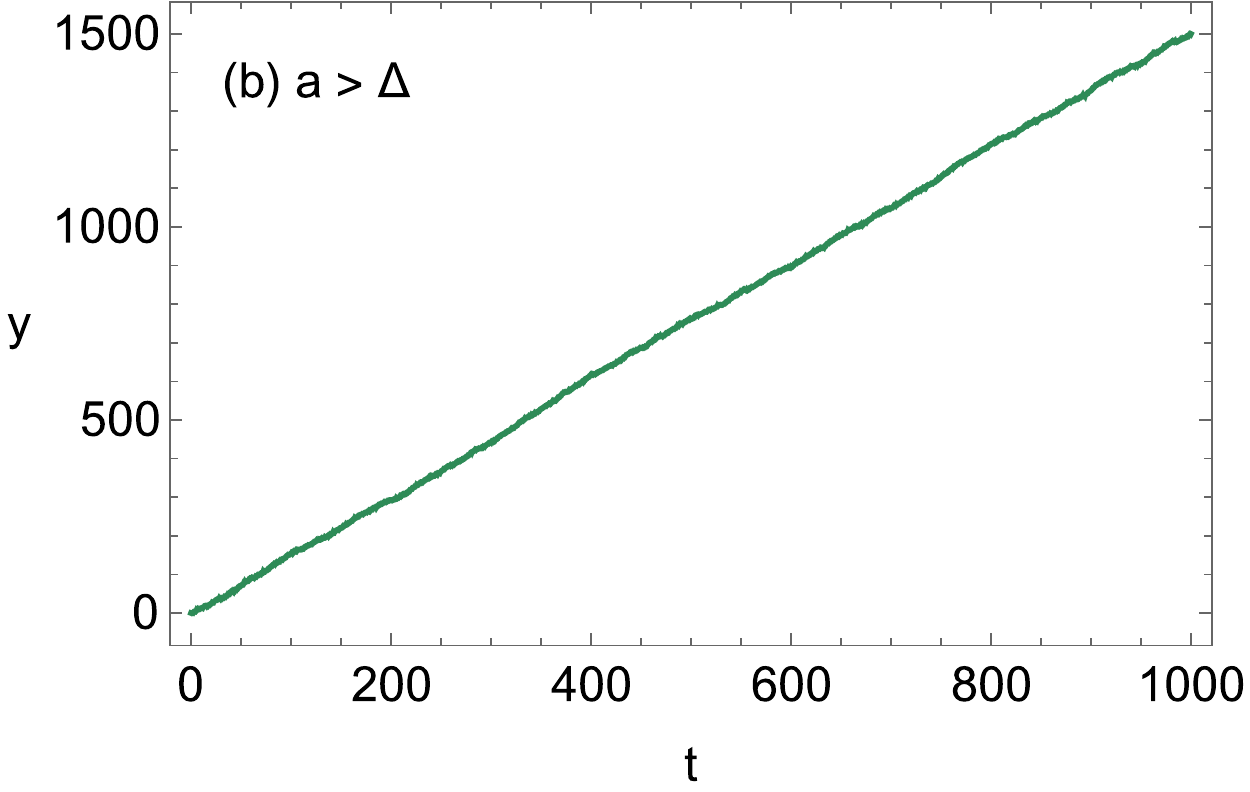}
\caption{A sample trajectory of the Brownian bridge with dry friction starting at $y(0)=0$ and ending at $y(T)=aT$ for (a) $a=0.5\leq \Delta=1$ and (b) $a=1.5>\Delta=1.0$, calculated at discrete time intervals of size $\epsilon=0.2$; other parameters are $T=1000$ and $\sigma=1$. In (a), red line denotes the expected beginning of the ascent at $\tau=(1-a/\Delta)T$.}
\label{fig4}
\end{figure}

Particular realisations of these two cases $a <\Delta$ and $a > \Delta$ are presented in figures \ref{fig4}(a) and \ref{fig4}(b) respectively.

%
%

\begin{figure}[hbt]
\centering\includegraphics[width=8cm]{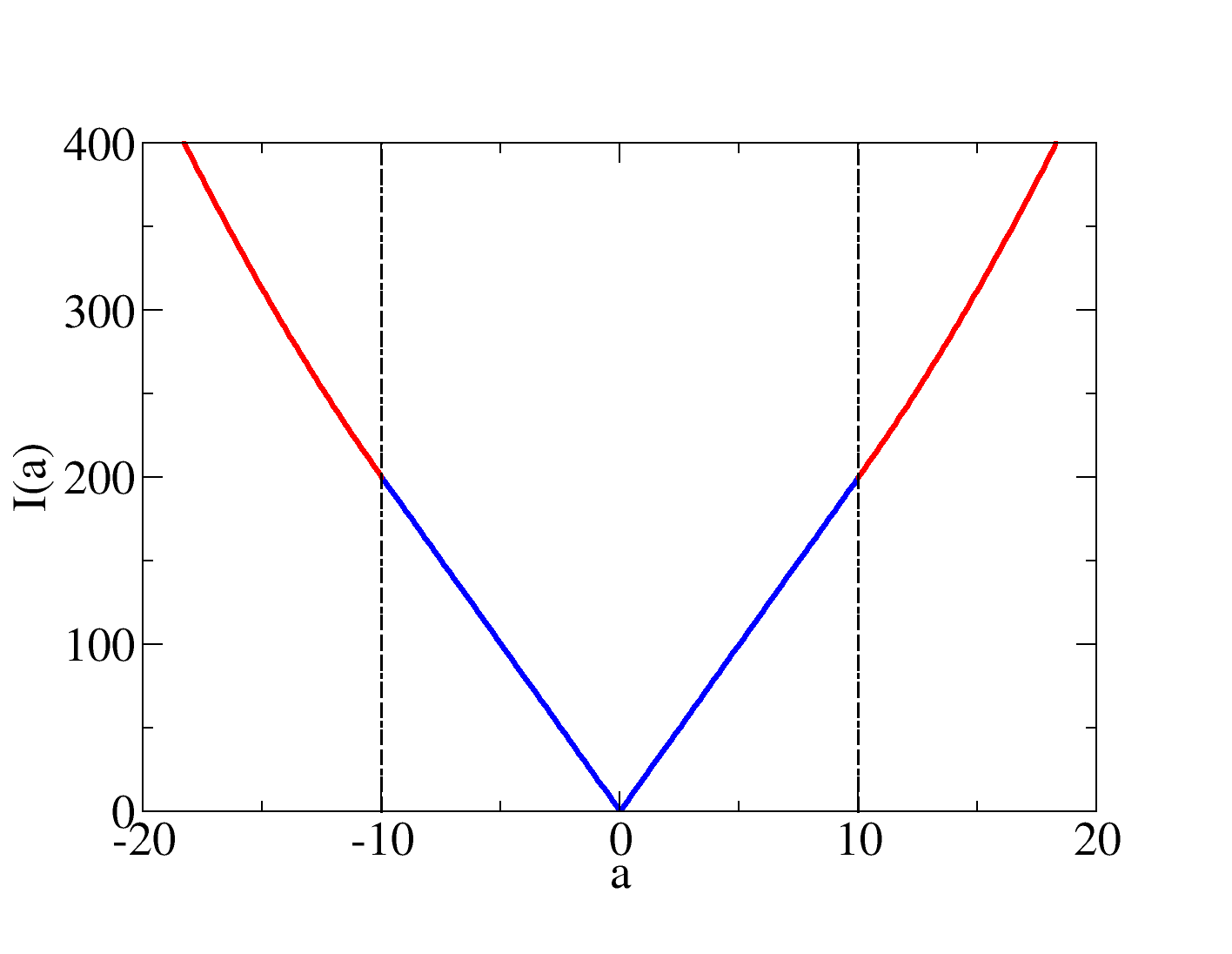}
\caption{Rate function $I(a)$ describing large deviations of $P(x(T)/T=a)\propto\textrm{exp}(-TI(a))$ for Brownian motion with dry friction; other parameters are $\Delta=10$ and $\sigma=1$. Dashed lines separate linear dependence on $a$ (red lines) where the equivalence of nonequilibrium path ensembles is partial, from quadratic dependence (blue lines) where the equivalence is full.}
\label{fig5}
\end{figure}

Despite the fact that the constrained process for $a<\Delta$ behaves in the same way as the unconstrained process for a macroscopic fraction of time $T$, we can check that the large deviation principle (\ref{ld}) holds. Indeed, from the solution of the Fokker-Planck equation in (\ref{fp_solution_dry}) we find that $P(A_T/T=a)$ satisfies the large deviation principle in (\ref{ld}) with rate function
\begin{equation}
I(a)=\cases{\frac{2\Delta \vert a\vert}{\sigma^2},& $\vert a\vert\leq\Delta$\cr \frac{(\vert a\vert+\Delta)^2}{2\sigma^2},& $\vert a\vert>\Delta$.\cr}
\end{equation}
\noindent One can also easily show that $I(a)$ is differentiable everywhere except at $a=0$, so that the tilting parameter $s$ is given by $s=I'(a)$ for $a\neq 0$. However, because $I(a)$ is linear for $\vert a\vert\leq\Delta$ (see Figure \ref{fig5}), the tilting parameter $s=2\Delta/\sigma^2\textrm{sgn}(a)$ depends only on the sign of $a$ and not on its absolute value. In this situation, the driven process with the tilting parameter $s=2\Delta/\sigma^2\textrm{sgn}(a)$ corresponds to a {\em range} of constrained processes that all have $\vert a\vert < \Delta$: the role of $a$ is to parametrise the fraction of  time $\tau\approx(1-a/\Delta)T$ spent around the steady state of the unconstrained process before it typically starts to ascend with the drift $\Delta$\footnote{Technically speaking, this will be true only in the limit $T\rightarrow\infty$. For a large, but finite $T$, the straight line in $I(a)$ may have a non-linear correction that selects a particular $a$.}. This can be thought of as  phase coexistence between
a drift-free process and a process with drift $\Delta$,
both lasting  finite fractions of the duration $T$.
This phenomenon
is known as partial equivalence\footnote{We also note that the rate function $I(a)$ is not differentiable at the point $a=0$, which means that for $a=0$ the tilting parameter $s$ is not unique. This fact has also been referred to as partial equivalence in some works \cite{CasettiKastner07}.}, the hallmark of which is exactly this kind of phase coexistence \cite{EllisHavenTurkington00,Touchette11}.

%

Let us conclude the analysis of diffusion bridges in this Section by mentioning that the last three diffusion processes - the Ornstein-Uhlenbeck process, the CIR process and the Brownian motion with dry friction - are somewhat special, in the sense that these processes respect detailed balance. This fact can be then exploited to understand why the trajectories depicted in Figures \ref{fig2}, \ref{fig3} and \ref{fig4}(a) behave initially as if they are not affected by the conditioning $A_T=aT$. 

To this end, let us consider a diffusion process $x(t)$ on the interval $[0,T]$ that starts at $x(0)=aT$, and has a time-independent drift $b(x)$ and a time-independent diffusion coefficient $\sigma(x)$. Let us assume that $p(x,t)$ is solution of the corresponding Fokker-Planck equation with initial condition $p(x,0)=\delta(x-aT)$, so that the process $x(t)$ starts at $aT$ at $t=0$. The detailed balance condition states that
\begin{equation}
\label{detailedbalance}
p(x,T-t\vert y,0)p_s(y)=p(aT,T-t\vert x,0)p_s(x),
\end{equation}
\noindent where $p_s(x)$ is the corresponding stationary distribution,

\begin{equation}
p_s(x)=\frac{\mathcal{N}}{\sigma(x)^2}e^{2\int^{x}\rmd x'b(x')/\sigma(x')^2}
\end{equation}
\noindent and $\mathcal{N}$ is the normalisation constant.

Next, we consider the time-reversed process $\bar{x}(t)$, defined as $\bar{x}(t)=x(T-t)$. This process is known to be again a diffusion process, governed by the following (It\^{o}) stochastic differential equation \cite{HaussmannPardoux86,MilletNualartSanz89},
\begin{equation}
\frac{d\bar{x}}{dt}=\bar{b}(x,t)+\sigma(x)\eta, 
\end{equation}
\noindent where $\bar{b}(x,t)$ is given by
\begin{equation}
\label{timereverseddrift}
\bar{b}(x,t)=-b(x)+\frac{1}{p(x,T-t\vert aT,0)}\frac{\partial}{\partial x}[\sigma^2 p(x,T-t\vert aT,0)].
\end{equation}
\noindent Using the detailed balance condition in (\ref{detailedbalance}), we can rewrite $\bar{b}(x,t)$ in (\ref{timereverseddrift}) as 
\begin{equation}
\bar{b}(x,t)=-b(x)+\sigma^2\frac{\partial}{\partial x}\textrm{ln}p(aT,T-t\vert x,0)+\frac{\partial \sigma^2}{\partial x}+\sigma^2\frac{\partial}{\partial x}\textrm{ln}p_s(x).
\end{equation}
\noindent The last term can be written as $2b(x)-\partial \sigma^2/\partial x$ so that the final expression for $\bar{b}(x,t)$ is given by
\begin{equation}
\bar{b}(x,t)=b(x)+\sigma^2\frac{\partial }{\partial x}\textrm{ln}p(aT,T\vert x,t),
\end{equation}
\noindent where we used the fact that $p(x,t\vert x',t')=p(x,t-t'\vert x',0)$. Notice that $\bar{b}(x,t)$ is the same as the drift $c_T(x,t)$ derived previously for a diffusion bridge, see (\ref{drift_bridge}). We have thus shown that the process conditioned to reach $aT$ at time $t=T$ is in fact the time-reversed process of the one that starts from $aT$, the only difference being that the end point of the time-reversed process $\bar{x}(T)$ is unconstrained, rather than fixed as for the diffusion bridge\footnote{That is not so important here, because we used initial conditions that are also stationary points (attractors) of the noiseless equation.}. Starting from $aT$ at time $t=0$, the Ornstein-Uhlenbeck process and the Brownian motion with dry friction both relax to their stationary distributions, which explains why the trajectories depicted in Figure \ref{fig2} and \ref{fig3}(a) are initially not affected by the conditioning $x(T)=aT$. 

It is noteworthy that we can extend this result further to simulate any diffusion bridge derived from a process that respects detailed balance, but whose full time-dependent solution of the Fokker-Planck equation is not known explicitly. The idea, proposed in Ref. \cite{BladtSorensen14}, is to simulate two processes, one that starts from $x(0)=0$ and the other that runs backwards in time and starts from $\bar{x}(0)=aT$, until they intersect at some later time $t'$; if $0<t'<T$, we can construct a diffusion bridge $y(t)$ as a piecewise function defined by $y(t)=x(t)$ for $t<t'$ and $y(t)=\bar{x}(t)$ for $t'<t<T$
Finally, let us emphasise that the condensation phenomenon studied in this Section is by no means due to the detailed balance only. Rather, it is  caused by a cost of having a large value of $X_t$ for all $0<t<T$, that is associated with a confining potential $V(x)$ (defined such that $b(x)=-V'(x)$) that grows to infinity faster than linearly. For example, an overdamped Brownian particle in a periodic and bounded potential (with or without external driving that breaks detailed balance), conditioned on $X_T=aT$ will show no condensation phenomenon \cite{ChetriteTouchette14}.


\section{Random walk bridges}
\label{rw_bridges}

So far we have presented diffusion processes, in which the noise was always Gaussian. In this Section we analyse continuous-space random walks in which the noise is not necessarily Gaussian. Although such processes typically converge to diffusions, some rare events that violate condition A are not captured in this limit. In particular, we are interested in a simple random walk on a real line, defined by the following stochastic recurrence equation
\begin{equation}
\label{rw}
X_t=X_{t-1}+\eta_t,\quad X_0=0,
\end{equation}

\noindent where $\eta_t$, $t=1\dots,T$ are independent and identically distributed random variables with a common probability density $\varphi(\eta_t)$, not necessarily Gaussian; we assume that $\langle\eta_t\rangle=\mu$ and $\langle\eta_t^2\rangle-\langle \eta_t\rangle^2=\sigma^2$. As before, we are interested in the bridge process, obtained by conditioning $X_t$ on fixed value of $X_T=aT$, which amounts to fixing the value of the sum

\begin{equation}
X_T=\sum_{t=1}^{T}\eta_t=aT.
\end{equation}
\noindent The probability density $P[X\vert X_T=aT]$ of a path $X_t$ conditioned on $X_T=aT$ can be written as,

\begin{eqnarray}
\label{path_rw}
\fl\qquad P[X\vert X_T=aT]&=&\frac{1}{P(X_T=aT)}\prod_{t=1}^{T}p(X_{t}\vert X_{t-1})\delta\left(X_T-aT\right)\nonumber\\
&& \fl\qquad =\frac{1}{P(X_T=aT)}\prod_{t=1}^{T}\varphi(X_{t}-X_{t-1})\delta\left(X_T-aT\right)\nonumber\\
&& \fl\qquad =\frac{1}{P\left(\sum_{t=1}^{T}\eta_t=aT\right)}\prod_{t=1}^{T}\varphi(\eta_t)\delta\left(\sum_{t=1}^{T}\eta_t-aT\right)
\end{eqnarray}

\noindent We note that the final expression for $P[X\vert X_T=aT]$ has the same factorised form as the steady-state probability in the well-studied mass transport models (for a review, see \cite{EvansHanney05}), where $t$ denotes coordinate in space, and $\eta_t$ is the mass at the site $t$. A distinctive feature of these models is the phenomenon of real-space condensation, whereby for $a>\mu$ there is a single site that carries a macroscopic fraction $(a-\mu)T$ of the total mass $a T$, while the rest of the sites have a typical mass of $O(1)$ with mean $\mu$ \cite{MEZ05,EMZ06}; for a rigorous analysis, see \cite{GrosskinskySchutzSpohn03,CG10, ArmendarizLoulakis11}. This situation occurs only if the underlying steady-state weight $\varphi(\eta)$ is heavy-tailed \cite{Linnik61,Nagaev69}, in the sense that

\begin{equation}
\int \rmd\eta\varphi(\eta)e^{k\eta}=\infty\quad\textrm{for $k>0$},
\end{equation}

\noindent which is to say that the moment-generating function of $\eta$ does not exist. Examples of heavy tails are a stretched exponential $\varphi(\eta)\propto\exp(-a\eta^{\alpha})$ with $\alpha<1$ or a power law
$\varphi(\eta)\propto A/\eta^b$ with $b>2$. In all these cases, $P(X_T=aT)$ takes the following form in the limit of large $T$

\begin{equation}
P(X_T/T=a)=T\varphi(a-\mu), \quad T\rightarrow\infty,
\end{equation}

\noindent so that the large deviation principle (condition A) does not hold. The particular form of $P(X_T/T=a)$ tells us that the event $X_T/T=a$ is realised by a single random variable taking a large value; this random variable can be any of the $T$ random variables $\eta_t$, $t=1,\dots,T$, hence the prefactor $T$.
%
%
\begin{figure}[hbt]
\centering\includegraphics[height=5cm]{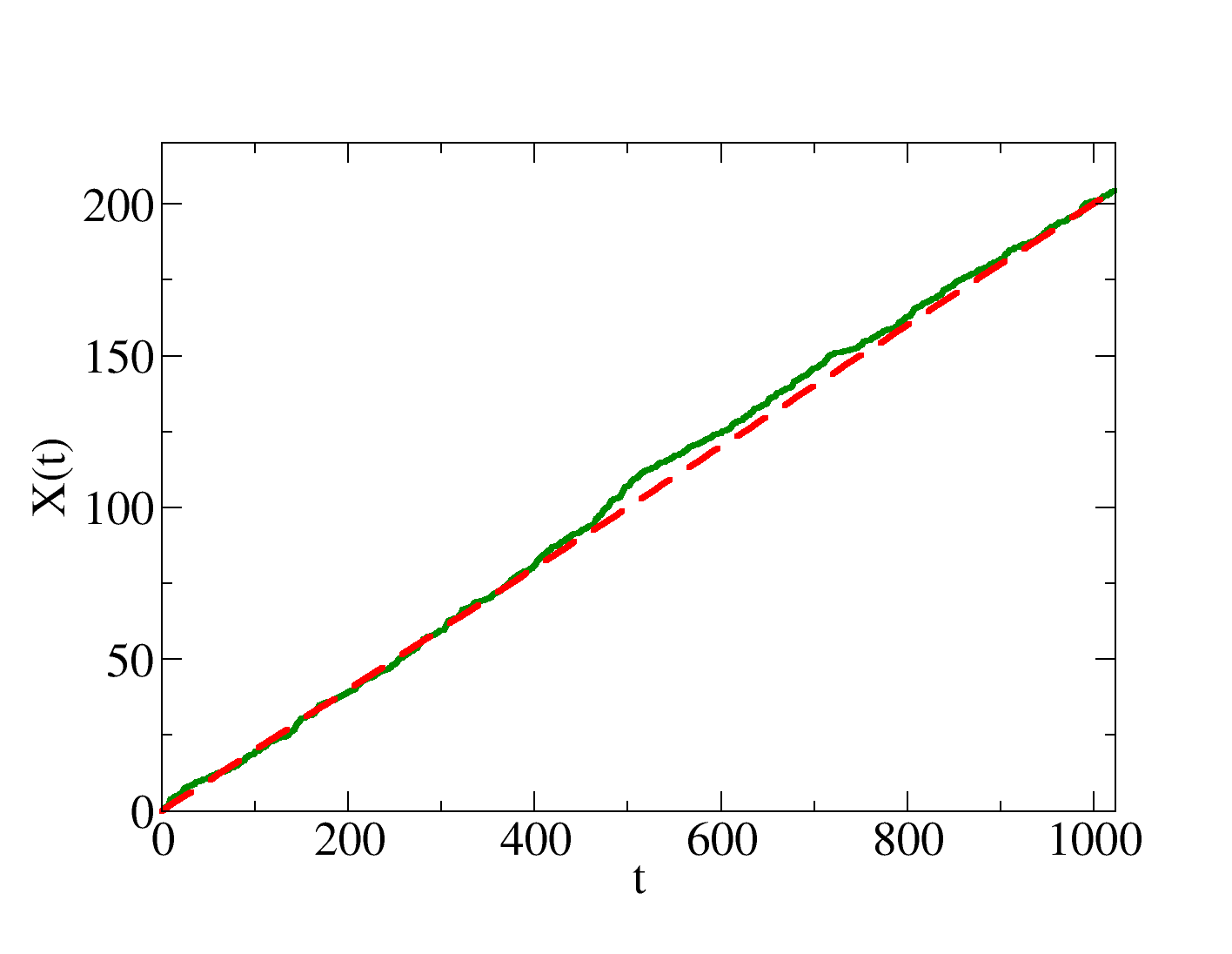}
\includegraphics[height=5cm]{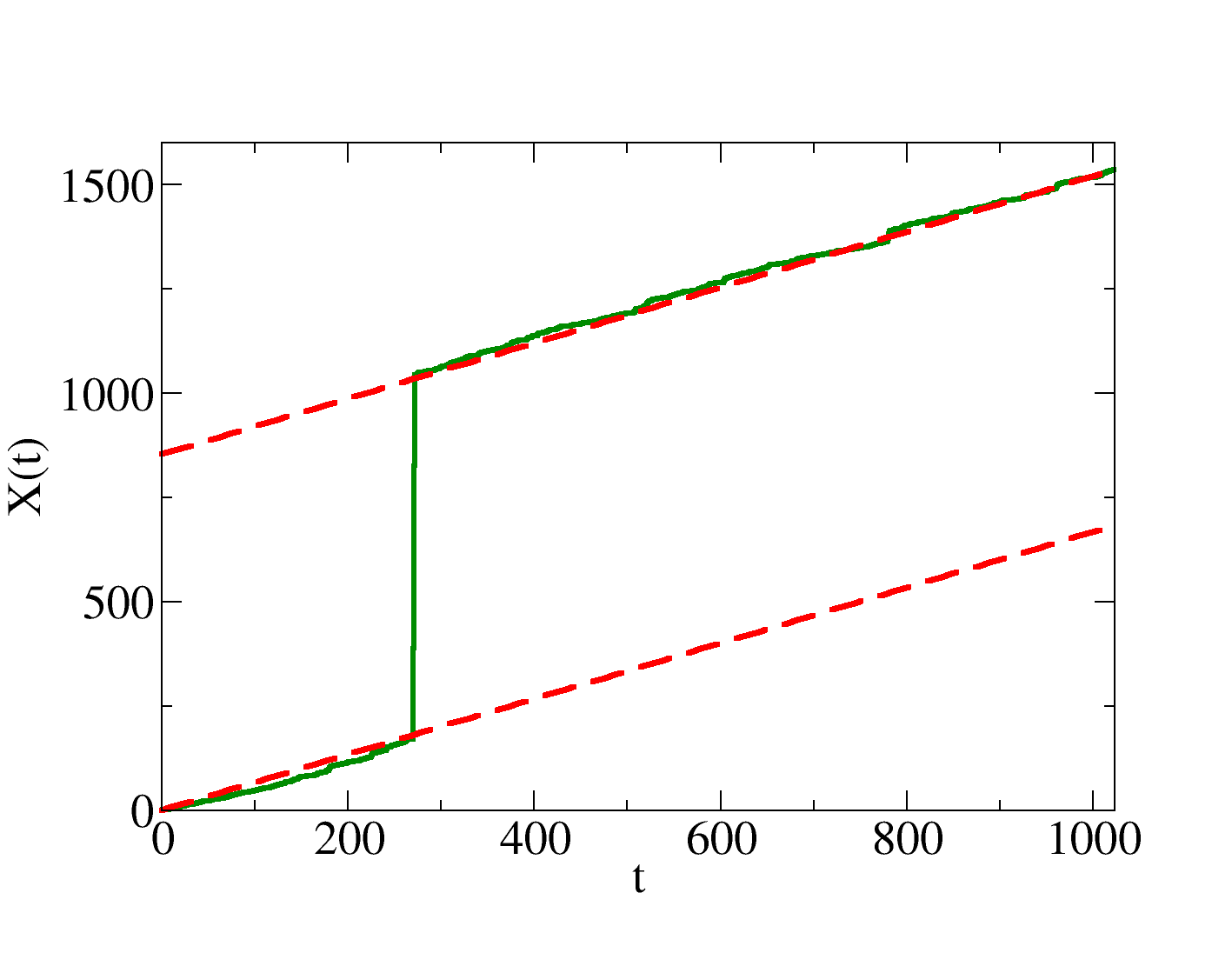}
\caption{Random walk constrained on the fixed value of $X_T=\sum_{t=1}^{T}\eta_t=aT$, for $T=1024$ and (a) $a=0.2<\mu$ and (b) $a=1.5>\mu$. Here $\varphi(x)=\alpha/(x+1)^{\alpha+1}$, where $\alpha=5/2$ (the Lomax distribution), which has the mean of $\mu=1/(\alpha-1)=2/3$. Dashed line in (a) has a slope of $a=0.2$, compared to dashed lines in (b) which have a slope of $\mu=2/3$; in (b), the size of the jump is $(a-\mu)T=5/6 T$.}
\label{fig6}
\end{figure}
An example of the random walk bridge (\ref{rw}) driven by a heavy-tailed noise is presented in Figure \ref{fig6} for (a) $a<\mu$ and (b) $a>\mu$, the latter showing a distinctive jump of size of $(a-\mu)T$

Recently, we reported a condensation transition for independent and identically distributed random variables $\eta_t$, $t=1,\dots,T$ constrained to have fixed values of both $M=\sum_{t=1}^{T}\eta_t\equiv \rho T$ and $V=\sum_{t=1}^{T}\eta_{t}^{1/p}=\delta T$ \cite{SNEM14a,SNEM14b}, where $p\neq 1$ is a parameter. In this situation, there is no need for the distribution of $\eta_t$ to be heavy-tailed - instead condensation is achieved through the second constraint, which changes the bare (light-tailed) probability density to a heavy-tailed one.
For $p<1$, the condensation transition happens for $\delta>\delta_c$, where $\delta_c$ is the mean of the effective heavy-tailed distribution, so that the value of $V$ falls in the large deviation regime.

Applied to the random walk in (\ref{rw}), the second constraint for $p<1$ takes the form of $A_T=\sum_{t=1}^{T}(X_{t}-X_{t-1})^{1/p}$; it is often called realised power variation, or realised quadratic variation specifically for $p=1/2$ \cite{BNS03,BNS04} and measures the variation of the trajectory. However, as we showed in Ref. \cite{SNEM14a,SNEM14b}, the condensation in this context may be viewed as a finite-size effect for $p<1$, in the sense that the joint probability density behaves as $P(M,V)\sim e^{-TI(\rho)+O(T^{\gamma p})}$, where the correction due to the condensation is of sublinear order in $L$; here $1\leq\gamma<1/p$ is related to the tail of the probability density in the case $\varphi(\eta)\sim e^{-k\eta^{\gamma}}$ for large $\eta$. This further means that the equivalence of path ensembles is actually restored in the limit $T\rightarrow\infty$, in which the size of the jump $T^{\gamma p}$ relative to $T$ goes to zero. On the other hand, for $p>1$ the jump is of  $O(T)$
and implies ensemble inequivalence in the limit $T\rightarrow\infty$.
%
%
\begin{figure}[hbt]
\centering\includegraphics[height=5cm]{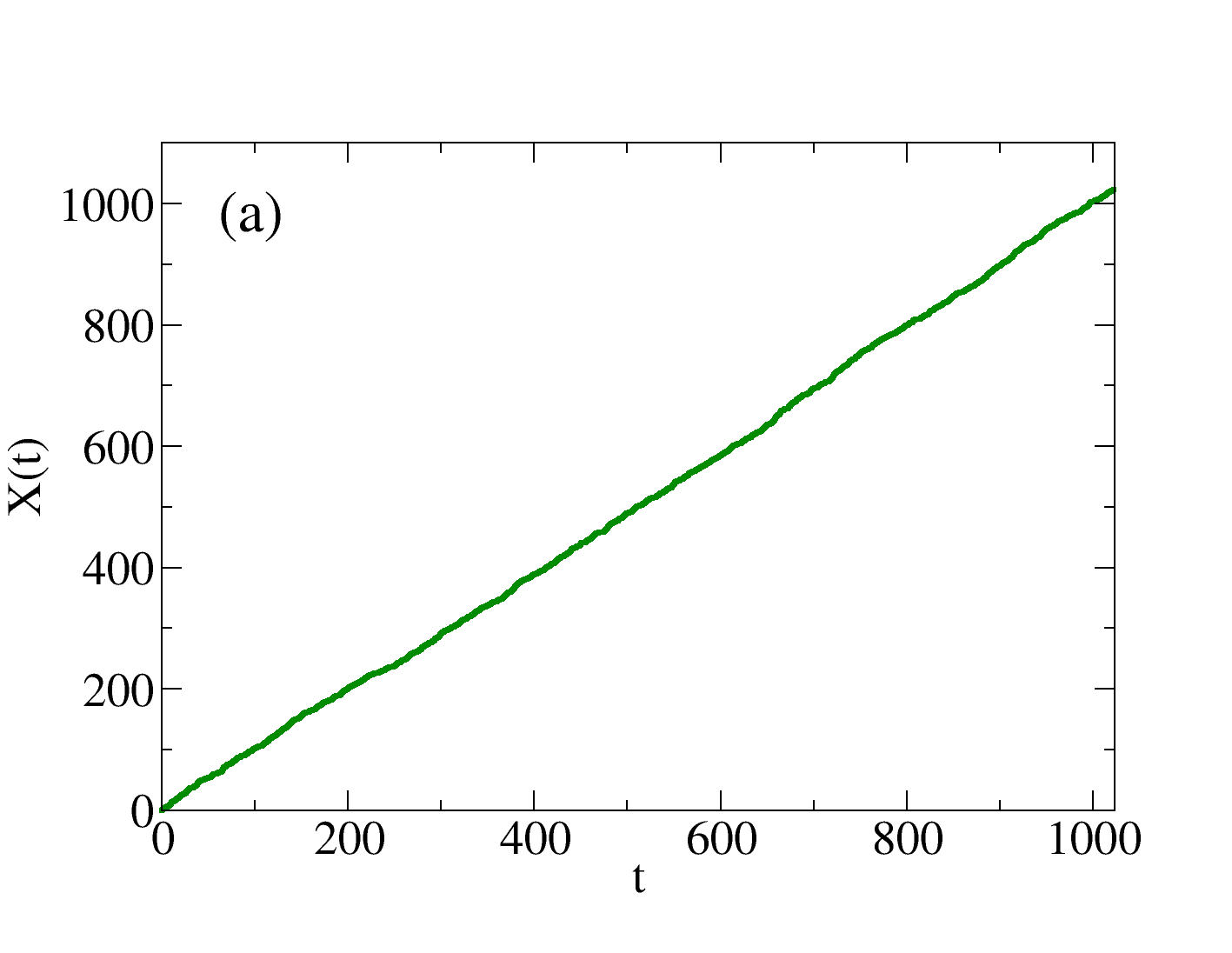}
\includegraphics[height=5cm]{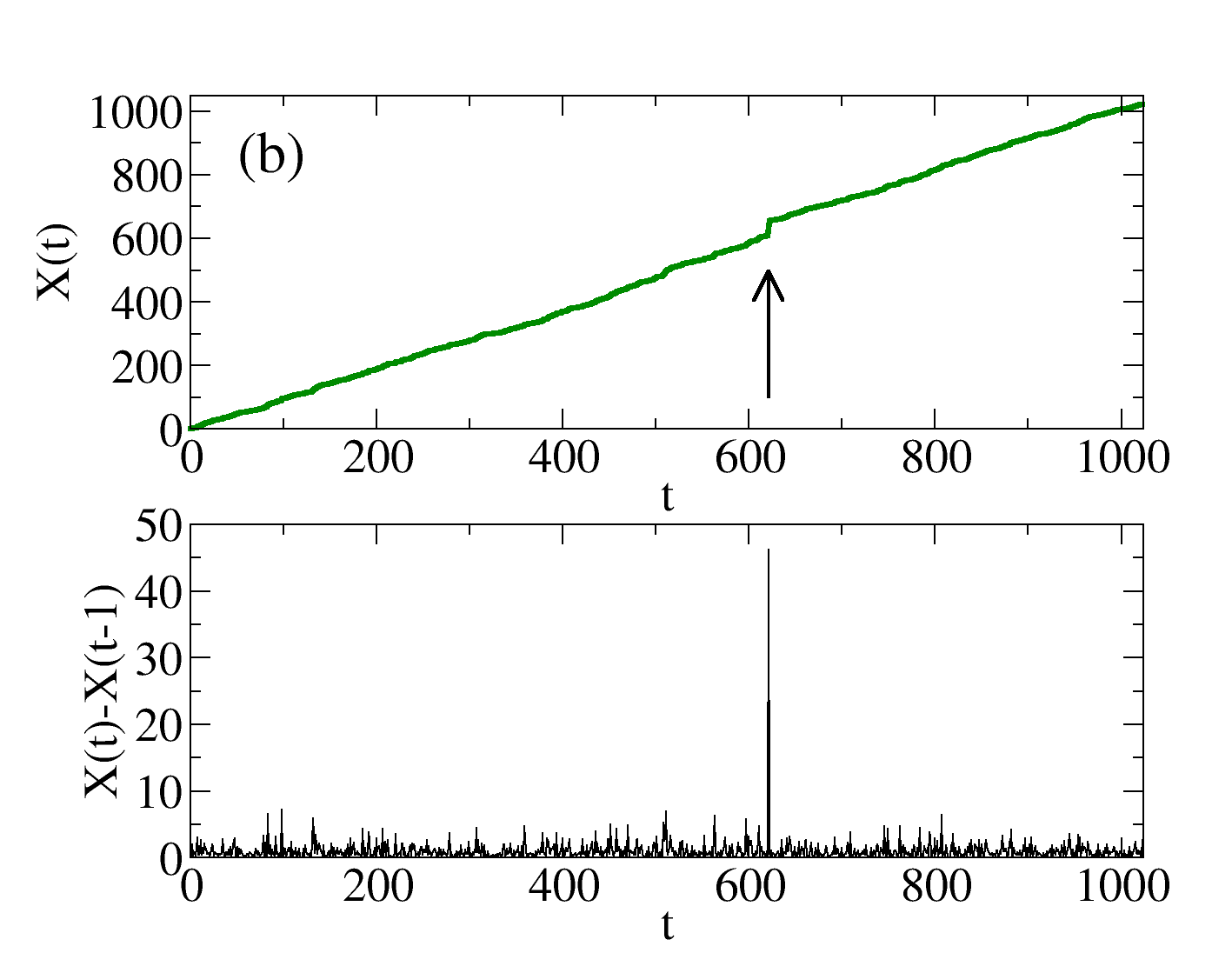}
\caption{Random walk constrained on the fixed values of $X_T=\sum_{t=1}^{T}\eta_t=aT$ and $V=\sum_{t=1}^{T}\eta_{t}^{2}=\delta T$, for $\mu=1$, $T=1024$ and (a) $\delta=3/2<\delta_c$ and (b) $\delta=4>\delta_c$; here $\eta_t$ is exponentially distributed for which $\delta_c$ can be computed analytically and reads $\delta_c=2\mu^2=2$. The second plot in (b) shows a condensate at time $t=621$ of size $\approx \sqrt{\delta-\delta_c}L$.}
\label{fig7}
\end{figure}

An example of a random walk constrained on the fixed values of $X_T=\sum_{t=1}^{T}\eta_t=aT$ and $V=\sum_{t=1}^{T}\eta_{t}^{2}=\delta T$ is presented in Figure \ref{fig7}, for (a) $\delta<\delta_c$ and (b) $\delta>\delta_c$, for an exponentially distributed noise $\eta_t$. While the condensate is visible in the noise variables $\eta_{t}$ (lower figure in Figure \ref{fig7}b), it is only a minor jump (that scales sublinearly with $T$) in the overall sample trajectory $X_t$.


\section{Conclusions}
\label{conclusions}
The s ensemble approach holds great promise for studying large fluctuations in nonequilibrium systems, regardless of whether they obey detailed balance or not. Its connection to the corresponding constrained dynamics leading to a large fluctuation is often taken for granted, and is expected to hold in majority of cases. The connection has been recently established rigorously, based on conditions rooted in the large deviation theory.

In this work, we presented several constrained stochastic systems in which one or two of these conditions are not met, and we studied whether they are equivalent to their corresponding driven processes in the s ensemble approach. For a one-dimensional stochastic variable $X_t$, we looked at large deviations of its time-integrated speed on the interval $[0,T]$, which amounts to conditioning on the fixed value of $X_T$. Such constrained stochastic processes, called stochastic bridges, are particularly convenient for the present study because their dynamics is Markovian and can be constructed exactly for the whole interval $[0,T]$, and not just in the limit of large $T$. As a main result, we showed several examples in which the constrained and driven dynamics are not equivalent in the limit of large $T$. Notably, this is manifested by condensation-like phenomena, in the sense that to meet the conditioning, the constrained process changes only a small portion of the dynamics. We have found essentially two types of condensation phenomena in these examples, both related to anomalous large deviations.

The first type of condensation is where a large deviation is not realised in the interior of the interval $[0,T]$, but is rather a boundary effect realised at time $T$. One such example, the Ornstein-Uhlenbeck bridge, was first reported in Ref. \cite{ChetriteTouchette14}. In the present study we analysed several other diffusion bridges and argued that a similar boundary effect leading to ensemble inequivalence is expected whenever a large deviation of $X_T$ is `penalised' by a confining potential. In addition, we presented a borderline case of Brownian motion with dry friction, where constrained dynamics is modified on a finite fraction of the interval that scales linearly with $T$, which presents an example of partial equivalence. The second type of condensation is where a large deviation is realised by a single random variable, and is due to heavy-tailed probability distributions; the same phenomenon (in space, rather than in time) has been observed in the  stationary states of mass-transport models such as the zero-range process. 

The presented examples are by no means an exhaustive list of processes that exhibit condensation and inequivalence of nonequilibrium path ensembles. In fact, the equivalence established in Refs. \cite{ChetriteTouchette13,ChetriteTouchette14,Touchette15} provides a powerful tool for  understanding other condensation phenomena whose mechanism of condensation is not obvious. Here we briefly mention the case of interaction-driven condensation reported in Ref. \cite{EvansHanneyMajumdar06}, in which the original zero-range process was generalised to include hopping rates that depend not only on the departure site, but also on its immediate environment, causing correlations between neighbouring random variables. The stationary probability of this process still admits a factorised form, which reads
\begin{equation}
\label{pairfactorized}
P(\{m_i\})\propto\prod_{i=1}^{L}w(m_{i},m_{i+1})\delta\left(\sum_{j=1}^{L}m_i-\rho L\right),
\end{equation}
\noindent where the delta function ensures that the total mass is conserved. In Ref. \cite{EvansHanneyMajumdar06}, the following choice for $w(m_{i},m_{i+1})$ was used
\begin{equation}
\label{w}
w(m,n)=\textrm{exp}\left[-J\vert m-n\vert\right]\cdot\textrm{exp}\left[\frac{1}{2}U\delta_{m,0}+\frac{1}{2}U\delta_{n,0}\right],
\end{equation}
\noindent where $\delta_{m,n}$ stands for the Kronecker delta function, and $J$ and $U$ are constants. When inserted into (\ref{pairfactorized}), the first factor in (\ref{w}) contributes to the probability of a random walk path, which becomes apparent by making the following change in notation: $i\rightarrow t$, $m_i\rightarrow X_t$ and $w(m_{i},m_{i+1})\rightarrow \varphi(X_{t+1}\vert X_t)$, where $\varphi$ is the same as in (\ref{path_rw}). The extra factor in (\ref{w}), when inserted in (\ref{pairfactorized}), appears in the path probability $P[X]$ as a factor of $\textrm{exp}(UTL_T)$, where $L_T$ is given by
\begin{equation}
\label{local_time}
L_T=\frac{1}{T}\sum_{t=1}^{T}\delta_{X_t,0}.
\end{equation}
\noindent The time-integrated observable $L_T$ measures the number of returns to the origin and is a discrete-time analogue of local time. The final expression for the path probability $P[X]$ thus has a hard constraint on the sum $\sum_{t=1}^{T}X_t$, which is the total area under the trajectory, and the tilting factor of $\textrm{exp}(UTL_T)$ related to conditioning on local time. The latter conditioning ensures that the number of visits to the origin scales with the total time $T$, so that the whole process is a series of random walk excursions (recurrent trajectories that stay positive), conditioned on a fixed value of the total area and on the total duration $T$. One can show that these two constraints are responsible for the condensation. Details of this calculation, which is similar to the one for the constraint-driven condensation \cite{SNEM14a,SNEM14b}, will be presented elsewhere.

\ack
JSN and MRE would like to acknowledge funding from EPSRC under grant number EP/J007404/1. JSN and MRE would like to thank Hugo Touchette and Bernard Derrida for illuminating discussions and J\'{e}r\^{o}me Michaud for discussing many mathematical questions.


\end{document}